\documentclass[onecolumn,preprintnumbers,amsmath,amssymb,aps,prd]{revtex4-1}
\usepackage{bm,tikz}
\usepackage{color}
\usepackage{tikz}
\usetikzlibrary{decorations.pathmorphing,decorations.markings}

\usepackage{graphicx}
\graphicspath{{./}}
\usepackage{epsfig}
\usepackage{multirow}
\usepackage{float}
\usepackage[utf8x]{inputenc}
\usepackage{amssymb,amsmath,appendix}
\usepackage{slashed}
\usepackage{subfigure}
\usepackage{array}
\usepackage{tabularray}
\usepackage[colorlinks=true,citecolor=blue,urlcolor=blue,linkcolor=blue]{hyperref}
\usepackage{ulem}

\begin{document}
\title{Effects of dark boson mediated feeble interaction between dark matter (DM) and quark matter on $f$-mode oscillation of DM admixed quark stars}

\author{O. P. Jyothilakshmi$^{1}$}
\email{op\_jyothilakshmi@cb.students.amrita.edu}

\author{Lakshmi J. Naik$^{1}$}
\email{jn\_lakshmi@cb.students.amrita.edu}

\author{Debashree Sen$^2$}
\email{debashreesen88@gmail.com}

\author{Atanu Guha$^3$}
\email{atanu@cnu.ac.kr}

\author{V. Sreekanth$^{1}$}
\email{v\_sreekanth@cb.amrita.edu}

\affiliation{$^{1}$Department of Physics, Amrita School of Physical Sciences, Amrita Vishwa Vidyapeetham, Coimbatore, India}
\affiliation{$^2$Center for Extreme Nuclear Matters (CENuM), Korea University, Seoul 02841, Korea}
\affiliation{$^3$Department of Physics, Chungnam National University,\\ 99, Daehak-ro, Yuseong-gu, Daejeon-34134, South Korea}

\date{\today}




\begin{abstract}

We investigate the behavior of the prominent non-radial fundamental $f$-mode oscillations of dark matter (DM) admixed strange quark stars (DMSQSs), by adopting an equation of state (EoS) developed in Ref.~\cite{Sen:2022pfr}, which considers the possible presence of feebly interacting DM in strange quark stars (SQSs) for the first time. Within the model, feeble interaction between fermionic DM $\chi$ and strange quark matter (SQM) is invoked via a vector new physics mediator $\xi$ with coupling strength $y_{\xi}$. The pure SQM is described by the vector MIT Bag model. By varying different EoS parameters, the structural properties viz. the mass, radius and tidal deformability ($\Lambda$) of the DMSQSs are studied with respect to various astrophysical constraints. We study in detail the $f$-mode spectra within the Cowling approximation by obtaining the frequencies as a function of mass, compactness and $\Lambda$ of the star. To the best of our knowledge, this study represents the first analysis of non-radial $f$-mode oscillations of DMSQSs. Our investigation indicates that the presence of DM and its interaction with SQM has great impact on the $f$-modes. We show that the $f$-mode frequencies are larger for DMSQSs, which are largely populated with massive DM fermions, compared to the SQSs. Further, we obtain a linear empirical relation between the $f$-modes and the average density of the star. We also find that the mass-scaled angular frequency varies universally with compactness and $\ln{\Lambda}$ for DMSQSs. Further, our studies indicate that the inclusion of DM in compact stars reduces the deviation of $f$-mode frequency from general relativistic to Cowling approximation.

\end{abstract}




\maketitle



\section{Introduction}
\label{Intro}

One of the most interesting facts about compact stars is that they present the opportunity to study the matter at extreme conditions of density (about $5-10$ times the normal nuclear density), compactness and gravity. Their composition at such high density is still experimentally inconclusive and presently we adopt various theoretical modeling of the constituent matter equation of state (EoS) to understand the interactions and structure of compact stars at such high density. However, the compact star EoS is constrained by several astrophysical and observational results for the structural properties like the mass, radius and tidal deformability. Such constraints include the data obtained from the most massive pulsar PSR J0740+6620~\cite{Fonseca:2021wxt, Miller:2021qha, Riley:2021pdl}, gravitational wave (GW170817)~\cite{LIGOScientific:2018cki}, 
PSR J0030+0451 (NICER)\cite{Riley:2019yda, Miller:2019cac}, and HESS J1731-347 \cite{Doroshenko2022}.

Many theoretical studies predict that strange quark stars (SQSs), composed of $u$, $d$ and $s$ quarks, may exist \cite{Olinto:1986je,Farhi:1984qu,Torres:2012xv,Miao:2021nuq,Li:2020wbw,Lugones:2022upj,Li:2019ztm,Li:2019akk,Luo:2019dpm,Pal:2023dlv}. Such prediction is based on the Bodmer-Witten conjecture, which states that the energy per baryon density $\varepsilon/\rho_B$ of strange quark matter (SQM) is lower than that of pure nucleonic system \cite{Bodmer:1971we,Chin:1979yb,Witten:1984rs,Farhi:1984qu}. Consequently, over many years, the theoretical modeling of SQM has been a topic of great interest. The first of such kind of models proposed is the well known MIT Bag model \cite{Chodos:1974je}, which is later modified to ensure the repulsive effect between the quarks by introducing a vector meson as mediator (vBag model) \cite{Klahn:2015mfa, Cierniak:2019hhe, Gomes:2018bpw, Franzon:2016urz,Wei:2018mxy,Lopes:2020btp,Kumar:2022byc}. In the MIT Bag model,  
the difference in energy density between the perturbative vacuum and the true vacuum is represented by a constant $B$ whose value is still not well known. The stability of SQS in terms of $\varepsilon/\rho_B$ is found to be controlled by $B$ \cite{Farhi:1984qu,Torres:2012xv} and further, the allowed range of $B$ for different vector meson couplings ($G_V$) is estimated in Ref.~\cite{Lopes:2020btp}. 
\par
Several astrophysical and cosmological observations like the rotation curves of the galaxies,  gravitational lensing, X-ray analysis of Bullet cluster \cite{Planck:2015fie,Bertone:2004pz,Aghanim:2018eyx,Bauer:2020zsj} support the 
fact that the dark matter (DM) to luminous baryonic matter ratio of the total energy budget in the Universe is around 5:1. The important experimental search avenues including the direct and the indirect detection strategies \cite{Lisanti:2016jxe,Profumo:2013yn} are actively pursuing the intricacies of interaction between the DM and standard model particles. Of these searches, the most important direct detection experiments are the SuperCDMS \cite{SuperCDMS:2018mne}, XENONnT \cite{XENON:2022ltv}, PandaX-II \cite{PandaX-II:2020oim}, DarkSide-50 \cite{DarkSide:2018ppu}, SENSEI \cite{Crisler:2018gci} and LUX-ZEPLIN~\cite{LZ:2022lsv}; while, 
the indirect searches are FERMI-LAT~\cite{Fermi-LAT:2009ihh}, IceCube~\cite{IceCube:2014stg}, PAMELA~\cite{PAMELA:2008gwm, PAMELA:2013vxg}, AMS-02~\cite{AMS:2014bun}, Voyager~\cite{Boudaud:2016mos} and CALET~\cite{CALET:2017uxd,Adriani:2018ktz}. However, the exact nature, properties and interaction of DM particle candidates are still largely unknown. The present day thermal relic abundances of DM has been prescribed to be $\sim \Omega_{DM} h^2 \approx 0.12$ \cite{ParticleDataGroup:2018ovx,Bauer:2017qwy,Cannoni:2015wba} from the Cosmic Microwave Background (CMB) anisotropy maps, obtained from the Wilkinson Microwave Anisotropy Probe (WMAP) data \cite{WMAP:2012fli}. Here, the relic abundances of DM ($\Omega_{DM}$) is conventionally estimated in terms of the critical density of the Universe $\rho_{crit}$ and the scale factor for the Hubble expansion rate $h$. Thus any reasonable DM model is constrained to successfully reproduce the observed non-baryonic relic density.
\par
The compact stars like SQSs are extremely gravitating objects which are capable of accreting matter from their surroundings and such accreted matter may include DM \cite{Baryakhtar:2017dbj,Joglekar:2019vzy,Bell:2019pyc,Joglekar:2020liw,Bramante:2023djs}. Few other phenomena may also be attributed to the possible presence of DM inside compact stars; like, dark decays of neutrons \cite{Husain:2022bxl,Husain:2022brl,Husain:2023fwb,Zhou:2023ndi}, inheritance of DM from the supernovae \cite{Nelson:2018xtr} etc. It is therefore reasonable to consider the DM particle density $\rho_{\chi}$ to be almost constant throughout the radial profile of the star \cite{Panotopoulos:2017idn,Guha:2021njn,Sen:2021wev,Sen:2022pfr,Guha:2024pnn}. The compact objects with DM admixture may also be studied within the two fluid approach and several works are there in this direction~\cite{Leung:2012vea,Rezaei:2016zje,Mukhopadhyay:2016dsg,Das:2020ecp,Mariani:2023wtv,Barbat:2024yvi,Zhen:2024xjc,Thakur:2023aqm}, though the interaction between DM and nuclear matter is not considered.

DM thus becomes structural part of the SQSs, consequently allowing us to study the properties of DM admixed SQSs (DMSQSs) in the light of the different observational and astrophysical constraints on the structural properties of compact stars \cite{Yang:2024sxi,Rezaei:2023iif,Panotopoulos:2017eig,Zheng:2016ygg,Mukhopadhyay:2015xhs,Jimenez:2021nmr,Yang:2021bpe,MafaTakisa:2020avv,Panotopoulos:2018ipq,Panotopoulos:2018joc,Cassing:2022tnn,Ferreira:2022fjo}. DM may or may not interact with SQM; however, if they interact, then the interaction strength must be extremely weak in order to prevent the star from collapsing \cite{Zheng:2016ygg}. Based on this notion, in Ref. \cite{Sen:2022pfr}, a feeble interaction between SQM and fermionic DM $\chi$ via a dark vector boson mediator $\xi$ was introduced. The pure SQM is described with the vBag model. The mass of DM fermion $m_{\chi}$ and the mass of vector mediator $m_{\xi}$ are chosen by following the self-interaction constraint $\sigma_T/m_\chi \leq 1.25~\rm{cm^2/g}$ \cite{Randall:2007ph, Robertson:2016xjh} from the Bullet cluster \cite{Tulin:2013teo, Tulin:2017ara,Hambye:2019tjt}. Here, $\sigma_T$ is the self-scattering transfer cross-section. Further, the coupling $y_\xi$ between $\chi$ and $\xi$ are obtained by satisfying the present day relic abundance of DM fermion $\chi$ \cite{Belanger:2013oya,Gondolo:1990dk,Guha:2018mli}. In the present work, we follow the same treatment as in Ref.~\cite{Sen:2022pfr} for the interaction between DM and SQM, in order to investigate its effects on the non-radial oscillations of the DMSQSs. 
\par 
Gravitational waves (GWs) can be produced by binary neutron star (BNS) mergers as well as from isolated neutron stars (NSs). Following the first detection of a GW signal from a binary black hole merger \cite{LIGOScientific:2016aoc}, many more signals have been detected from supernovae, NS mergers and supermassive black hole mergers. Furthermore, the NanoGrav collaboration recently revealed the existence of a cosmic gravitational wave background ~\cite{NANOGrav:2023gor,EPTA:2023fyk,Reardon:2023gzh,Xu:2023wog}. The detection of GWs from a BNS merger (GW170817)~\cite{LIGOScientific:2017vwq} by the advanced LIGO and VIRGO collaborations combined with the multi-messenger observations of BNS systems have paved a new way for studying the interior composition of compact stars. The non-radial pulsations of compact stars are of current interest in the era of gravitational wave detection because they are related to the emission of GWs of different frequencies viz., fundamental ($f$), pressure ($p$), rotational ($r$), space-time ($w$) and gravity ($g$) modes \cite{Kokkotas:1999bd}. There have been a lot of studies on these oscillation modes done over the past few years \cite{Lindblom:2001hd,Jha:2010an,Flores:2013yqa,Andersson:2019mxp,Jyothilakshmi:2022hys,Zhao:2022tcw,Kumar:2024jky,Guha:2024gfe}, among which $f$-modes have drawn a great interest since their frequencies ($1-3$ kHz) lie in the range of the future GW detectors such as the Einstein Telescope and the Cosmic Explorer and also the LIGO O4 run. With these next generation GW detectors the $f$-mode oscillations are expected to be observed; which will provide more powerful insights to the composition and interaction of matter inside compact stars and help in constraining the EoS in stringent manner. 
\par
For the purpose of gravitational wave asteroseismology, the estimation of the gravitational-radiation flux from compact stars is based on the theory of non-radial oscillations, which was initially developed in the framework of general relativity in \cite{1967ApJ...149..591T,Lindblom:1983ps}. Later, with the help of Cowling approximations \cite{Cowling:1941nqk}, the derivation of the integrated solution was simplified \cite{Sotani:2010mx} by neglecting the metric perturbations. 
Although, this approximation can result in overestimation of the $f$-modes upto 30\% \cite{Zhao:2022tcw} compared to a fully general relativistic treatment preferred for a precise quantitative analysis; Cowling approximation is widely used to study the characteristics of $f$-modes. In the present work, we calculated the deviation of the $f$-mode frequencies obtained using the Cowling approximation from the fully relativistic approach for neutron star (NS), dark matter admixed neutron star (DMNS) and SQS. We found that the deviation percentage of DMNS is less compared to that of NS and SQS, indicating that inclusion of DM may lower the deviation. Thus, we believe that using the Cowling approximation in our analysis of DMSQSs may not result in a higher deviation percentage.

Recently, there have been several studies on non-radial $f$-mode oscillations using the Cowling approximation for different stellar composition \cite{Sotani:2010mx,Flores:2013yqa,Ranea-Sandoval:2018bgu,VasquezFlores:2019eht,Pradhan:2022vdf,Kalita:2023rbz,Zhang:2023zth,Jyothilakshmi:2024zqn,Jyothilakshmi:2025wru,Jyothilakshmi:2025zps}. Further, the non-radial $f$-mode oscillations of quark stars~\cite{1990ApJ...363..603C,Kojima:2002iv,Sotani:2003zc} and dark matter admixed NSs \cite{Das:2021dru, Hong:2023udv,Shirke:2024ymc,Sen:2024yim} 
were also studied. In the case of DMSQSs, although the radial oscillations have attracted the interest of the community recently \cite{Panotopoulos:2017eig,Panotopoulos:2018ipq,Jimenez:2021nmr,Zhen:2024xjc}; the non-radial oscillations have not been explored so far. This sets the motivation for our analysis. 

In the present work, we are particularly interested to show how the presence of DM and its interaction with SQM affects the non-radial $f$-mode frequency of the oscillation of the DMSQSs using the Cowling approximation.

We arrange the present work as follows. In the next Sec. \ref{Formalism}, we briefly discuss the mechanism of invoking feeble interaction of DM with SQM via the new physics vector mediator $\xi$. We also discuss the methodology of estimating the structural properties like mass, radius, tidal deformability and the $f$-mode oscillation of the DMSQSs. We then present the results and corresponding discussions in Sec. \ref{Results}. We summarize and conclude in the final section \ref{Conclusion} of the paper. Throughout this paper we set $c=G=1$, where $c$ is the speed of light in vacuum and $G$ is the universal gravitational constant. 


\section{Formalism}
\label{Formalism}

\subsection{Dark matter admixed quark star model}
\label{Models}

Here, we review the DMSQS model as derived in Ref.~\cite{Sen:2022pfr}. The model consists of two main sectors: strange quark matter (SQM) and dark matter (DM), together with their interactions. The vector MIT Bag (vBag) model is employed to describe the $u$, $d$ and $s$ quarks along with the electrons $e$~\cite{Lopes:2020btp,Kumar:2022byc}.
The complete modified Lagrangian for such a system is given as \cite{Sen:2022pfr} 
\begin{eqnarray}
\mathcal{L} &=& \sum_f \bigg[\overline{\psi}_f \bigg(i \gamma_{\mu} \partial^{\mu} -g_{\xi}\gamma_{\mu}\xi^{\mu} - m_f \bigg) \psi_f - B \bigg] \Theta (\overline{\psi}_f \psi_f) \nonumber \\  
&-& \sum_f g_{qqV} \bigg[\overline{\psi}_f \bigg(\gamma_{\mu}V^{\mu}\bigg) \psi_f \bigg] \Theta (\overline{\psi}_f \psi_f) 
- \frac{1}{4} V_{\mu\nu} V^{\mu\nu} + \frac{1}{2} m_V^2 V_{\mu} V^{\mu} \nonumber \\  
&+& \overline{\psi}_l \bigg(i \gamma_{\mu} \partial^{\mu} - m_l\bigg) \psi_l - \frac{1}{4} F_{\mu\nu} F^{\mu\nu} + \frac{1}{2} m_{\xi}^2 \xi_{\mu} \xi^{\mu} \nonumber \\ &+& \overline{\chi}\bigg[i \gamma_{\mu} \partial^{\mu} -y_{\xi} \gamma_{\mu} \xi^{\mu} \bigg]\chi;
\protect\label{Eq:Lagrangian}
\end{eqnarray}
where $f\equiv$ ($u$, $d$, $s$) and the lepton $l\equiv e$. The term $ \overline{\psi}_f ( g_{\xi}\gamma_{\mu}\xi^{\mu} ) \psi_f$ represents the interaction between the quark fields $\psi_f$ and the new physics vector dark mediator $\xi$ of mass $m_\xi$, with a very feeble coupling strength $g_\xi$. The Bag constant is denoted by $B$ and the Heaviside step function $\Theta$ in the Lagrangian ensures that the quarks are confined within the bag. The first term in second line denotes the interaction between quarks,  mediated by the repulsive vector meson $V$ having mass $m_V$, with coupling strength $g_{qqV}$. Further, a universal coupling scheme is considered {\it i.e.}, $g_{uuV}=g_{ddV}=g_{ssV}=g_{qqV}$ and $G_V=(g_{qqV}/m_V)^2$ denotes the scaled coupling. Note that $G_V = 0$ results in the original form of the MIT Bag model without interactions. The kinetic term of the vector meson in the SQM is given by $\frac{1}{4}V_{\mu\nu} V^{\mu\nu}$ with $V_{\mu\nu}=\partial_\mu V_\nu - \partial_\nu V_\mu$. Further, the kinetic term of the dark mediator $\xi$ is given as $\frac{1}{4}F_{\mu\nu} F^{\mu\nu}$, where $F_{\mu\nu} = \partial_\mu \xi_\nu - \partial_\nu \xi_\mu$. The last line of Eq.~\eqref{Eq:Lagrangian} denotes the DM self interaction of fermionic $\chi$ involving bosonic dark mediator $\xi$, with coupling $y_\xi$. The interaction between the quarks $\psi_f$ and fermionic DM $\chi$, mediated by the vector dark boson $\xi$, can be understood with the help of the Feynman diagram (Fig. \ref{fig:feyman-diag}), which depicts the reversible process of $\chi
\overline{\chi} \stackrel{\xi}{\rightleftharpoons} {\psi}_f \overline{\psi}_f$.

\tikzset{
photon/.style={decorate, decoration={snake,amplitude=2pt, segment length=5pt}, draw=black},
particle/.style={draw=black, postaction={decorate}, decoration={markings,mark=at position .5 with {\arrow[draw=black]{>}}}},
antiparticle/.style={draw=black, postaction={decorate}, decoration={markings,mark=at position .5 with {\arrow[draw=black]{<}}}}
,gluon/.style={decorate, draw=black, decoration={coil,amplitude=4pt, segment length=5pt}}
}
\begin{figure}
\begin{center}
 \begin{tikzpicture}[thick,scale=1.0]
\draw[particle] (-1.5,1) -- node[black,above,sloped] {$ \psi_f $} (0,0);
\draw[antiparticle] (-1.5,-1) -- node[black,above,sloped] {$ \overline{\psi}_f  $} (0,0);
\draw[photon] (0,0) -- node[black,below,sloped] {$ \xi  $} (2,0);
\draw[particle] (2,0) -- node[black,above,sloped] {$  \chi $} (3.5,1);
\draw[antiparticle] (2,0) -- node[black,above,sloped] {$ \overline{\chi}  $} (3.5,-1);
\filldraw[black] (0,0) circle (1.5pt) node[anchor=south]{$g_\xi$};
\filldraw[black] (2,0) circle (1.5pt) node[anchor=south]{$y_\xi$};
\end{tikzpicture}
\hspace{1.5cm}
\begin{tikzpicture}[thick,scale=1.0]
\draw[particle] (-1.5,1) -- node[black,above,sloped] {$ \chi $} (0,0);
\draw[antiparticle] (-1.5,-1) -- node[black,above,sloped] {$ \overline{\chi}  $} (0,0);
\draw[photon] (0,0) -- node[black,below,sloped] {$ \xi  $} (2,0);
\draw[particle] (2,0) -- node[black,above,sloped] {$ \psi_f  $} (3.5,1);
\draw[antiparticle] (2,0) -- node[black,above,sloped] {$ \overline{\psi}_f  $} (3.5,-1);
\filldraw[black] (0,0) circle (1.5pt) node[anchor=south]{$y_\xi$};
\filldraw[black] (2,0) circle (1.5pt) node[anchor=south]{$g_\xi$};
\end{tikzpicture}
\end{center}
\caption{\,Feynman diagrams for the fermionic dark matter interaction with quarks via dark mediator, $ \chi
\overline{\chi} \stackrel{\xi}{\rightleftharpoons} {\psi}_f \overline{\psi}_f $.}
\label{fig:feyman-diag}
\end{figure}
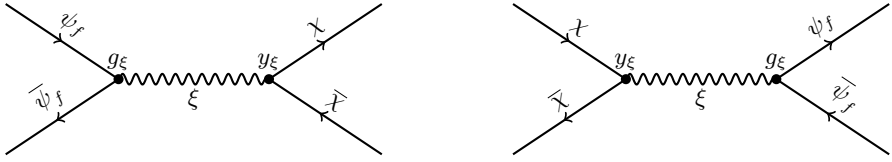

In this model, the stability window for Bag constant $B$ is fixed for a particular value of $G_V$, following the procedure described in Refs.~\cite{Torres:2012xv,Lopes:2020btp}. The upper bound of $B$ ($B_{max}$) is obtained from the Bodmer-Witten conjecture~\cite{Bodmer:1971we}, which demands that the energy per baryon density $\varepsilon/\rho_B$ at the surface of the SQS is less than the binding energy of iron nucleus {\it i.e.}, $\varepsilon/\rho_B \leq  930$ MeV~\cite{Torres:2012xv,Lopes:2020btp}; while the lower bound $B_{min}$ is obtained with two flavor QM. To obtain the EoS from Eq.~\eqref{Eq:Lagrangian}, we obtain an average value for $B$ ($B_{av}$) between $B_{max}$ and $B_{min}$, for a fixed value of $G_V$. Here, we take $G_V = 0.3, 0.5$, and 0.7 fm$^{2}$ and the corresponding values of $B$ have been shown in Table~\ref{Table:QM}. Further, we take the masses of quarks to be $m_s = 95$ MeV, $m_u = 5$ MeV and $m_d$ = 7 MeV and mass of vector meson is $m_V = 783$ MeV.

\begin{table}[h]
    \begin{tblr}{hlines, vlines,
            colspec = {Q[c,m,wd=1.6cm] Q[c,m,wd=1cm] Q[c,m,wd=1cm] Q[c,m,wd=1cm]},
            rowsep=1pt
            }
    $G_V$ (fm$^2$)   & 0.3   &   0.5   &    0.7 \\
    $B_{min}^{1/4}$ (MeV)  & 138  &  134   &  131  \\ 
    $B_{min}^{1/4}$ (MeV)  & 148  &  143   &  139  \\
    \end{tblr}
    \caption{Stability window for Bag constant $B$ obtained for the vBag model, for different values of $G_V$~\cite{Sen:2022pfr}.}
    \label{Table:QM}
\end{table}

It is interesting to note that, one of the most important features of the particular DMSQS model \cite{Sen:2022pfr} is that the DM parameters $m_\chi$, $m_\xi$, and $y_\xi$ are not independent of each other or arbitrary. The masses $m_\chi$ and $m_\xi$ are related by the self-interaction constraint from Bullet cluster~\cite{Tulin:2013teo, Tulin:2017ara,Hambye:2019tjt}. One can also find the graphical representation of the $m_\chi$ vs $m_\xi$ relation in Ref.~\cite{Sen:2022pfr}. The relation between $m_\chi$ and $y_\xi$ is obtained from the present day relic abundance of DM~\cite{Belanger:2013oya,Gondolo:1990dk,Guha:2018mli}. Hence, a variation in $m_{\chi}$ implies corresponding variation of both $m_{\xi}$ and $y_{\xi}$. We tabulate the mass of fermionic DM $\chi$ and the corresponding values of mass of the dark mediator $m_\xi$ and interaction strength $y_\xi$ used in the present analysis in Table.~\ref{Table:DM}. Note that the interaction strength between the dark mediator and quarks is considered to be feeble, $g_\xi \sim 10^{-4}$. Choosing the $g_\xi$ values less than $10^{-4}$ will not bring any further change to the EoS as well as the structural properties of DMSQS. Also, the values of $g_\xi > 10^{-4}$ are unphysical because in that case the interaction cross-section will be in
the ruled out region from direct detection experiments.
\begin{table}[]
    \begin{tblr}{hlines, vlines,
            colspec = {Q[c,m,wd=1.6cm] Q[c,m,wd=1cm] Q[c,m,wd=1cm] Q[c,m,wd=1cm] Q[c,m,wd=1cm] Q[c,m,wd=1cm] Q[c,m,wd=1cm]},
            rowsep=2pt
            }
    $m_\chi$ (GeV)   & 25  &  50  & 75  & 100 & 150  & 200 \\
    $m_\xi$ (MeV)    & 50  &  60  & 20  &  10 &  6    & 4  \\ 
    $y_\xi$        & 0.27 & 0.32 & 0.40 & 0.46 & 0.50 & 0.52  \\
    \end{tblr}
    \caption{Values of the mass of fermionic DM $m_\chi$ and the corresponding values of mass of dark mediator $m_\xi$ and interaction strength $y_\xi$~\cite{Sen:2022pfr}.}
    \label{Table:DM}
\end{table} 
\par 
In the mean field approximation, the equation of motion of the vector field in terms of $V_0$ is given by
\begin{eqnarray}
V_0=\frac{g_{qqV}}{m_V^2}\rho;
\end{eqnarray}
where, $\rho$ is the total quark density 
\begin{eqnarray}
\rho=\langle \psi_f^\dagger \psi_f\rangle=(\rho_u + \rho_d + \rho_s)=\frac{\gamma_f}{6 \pi^2} \sum_f {{k_F^3}_f}.
\label{rho}
\end{eqnarray}
Here, ${k_F}_f$ denotes the Fermi momenta of the quarks and $\gamma_f=6$ is the degeneracy factor for quarks. The vacuum expectation value ($\xi_0$) of the vector dark boson is
\begin{eqnarray}
\xi_0=\frac{g_{\xi}\rho + y_{\xi}\rho_{\chi}}{m_{\xi}^2};
\end{eqnarray}
where, the DM density is 
\begin{eqnarray}
\rho_{\chi}=\frac{\gamma_f}{6 \pi^2}{{{k_F^3}}_\chi}.
\label{rho_chi}
\end{eqnarray}
The number density $\rho_{\chi}$ is assumed to be constant throughout the radial profile of the star.
We note that this constant number density of fermionic DM is $\sim 1000$ times smaller than the average baryon number density of the SQM \cite{Panotopoulos:2017idn,Guha:2021njn,Sen:2021wev,Sen:2022pfr,Guha:2024pnn}. Thus the Fermi momentum of the DM fermions, ${k_F}_{\chi}$, is also constant over the radial profile of star. Unlike Ref.~\cite{Sen:2022pfr}, in the present analysis, we consider three different values for ${k_F}_{\chi}= 0.02, 0.03, 0.04$ GeV.
The quark chemical potential in presence of DM is modified as
\begin{eqnarray}
\mu_f=\sqrt{k_f^2 + m_f^2} + g_{qqV}V_0 + g_\xi \xi_0.
\end{eqnarray}
The calculated EoS for the DMSQSs is given as follows: 
\begin{eqnarray}
\varepsilon&=&\frac{1}{2}\frac{g_{qqV}^2}{m_V^2}\rho^2 + \frac{1}{2} \frac{\Big(g_{\xi}\rho + y_{\xi}\rho_{\chi}\Big)^2}{m_{\xi}^2} + \frac{\gamma_f}{2\pi^2}\sum_f \int_0^{{k_F}_f} \sqrt{k_f^2 + m_f^2}~ k_f^2~ dk_f \nonumber \\ 
&+& \frac{\gamma_l}{2\pi^2} \int_0^{{{k_F}}_l} \sqrt{k_l^2 + m_l^2}~ k_l^2~ dk_l + \frac{\gamma_\chi}{2\pi^2} \int_0^{{{k_F}}_\chi} \sqrt{k_\chi^2 + m_\chi^2}~ k_\chi^2~ dk_\chi + B, \\
\label{e}
P&=&\frac{1}{2}\frac{g_{qqV}^2}{m_V^2}\rho^2 + \frac{1}{2} \frac{\Big(g_{\xi}\rho + y_{\xi}\rho_{\chi}\Big)^2}{m_{\xi}^2} + \frac{\gamma_f}{6\pi^2}\sum_f \int_0^{{k_F}_f} \frac{k_f^4~ dk_f}{\sqrt{k_f^2+m_f^2}} \nonumber \\ 
&+&  \frac{\gamma_l}{6\pi^2} \int_0^{{{k_F}}_l} \frac{{{k_l^4}}~ dk_l}{\sqrt{k_l^2+m_l^2}} + \frac{\gamma_l}{6\pi^2} \int_0^{{{k_F}}_\chi} \frac{{{k_\chi^4}}~ dk_\chi}{\sqrt{k_\chi^2+m_\chi^2}} - B, \nonumber\\
\label{P}
\end{eqnarray} 
where $\varepsilon$ and $P$ are the energy density and pressure respectively. 
The presence of charge neutral DM do not affect the overall charge neutrality of the star and the EoS is computed by satisfying the charge neutrality and chemical potential equilibrium conditions of the star \cite{Glendenning:1997wn}.

\subsection{Structural properties and non-radial oscillations of dark matter admixed quark stars}
\label{Structure}

Using the EoS of DMSQSs, we estimate the global properties like the gravitational mass ($M$) and the radius ($R$) of the DMSQSs. 
In this section, we describe the formalism used to study the non-radial oscillations of the DMSQSs. The metric for a static spherically symmetric star is given as
\begin{eqnarray}
ds^2=-e^{2\Phi(r)}dt^2 + e^{2\lambda(r)}dr^2 + r^2d\theta^2 + r^2 \sin^2\theta d\phi^2.
\label{metric}
\end{eqnarray}
Here, $\Phi$ and $\lambda$ are the metric functions. The Einstein field equations are solved for the given metric for an ideal fluid to obtain the well-known Tolman-Oppenheimer-Volkoff (TOV) equations~\cite{Tolman:1939jz,Oppenheimer:1939ne} and is given as 
\begin{eqnarray}
\frac{dP(r)}{dr}&=&-\Big(\varepsilon(r)+P(r)\Big)\frac{d\Phi(r)}{dr},
\label{tov}\\
\frac{d\Phi(r)}{dr}&=&\frac{M(r)+4\pi r^3 P(r)}{r\Big(r-2 M(r)\Big)},
\label{tov2}\\
\frac{dM(r)}{dr}&=& 4\pi r^2 \varepsilon(r).
\label{tov3}
\end{eqnarray} 
These coupled differential equations are integrated from the center ($r=0$) to the surface ($r=R$) of the star. Towards the center of the star the pressure and mass of the star are $P_c=P(r=0)=P(\varepsilon_c)$ and $M_c=M(r=0)=0$, respectively. The pressure approaches zero towards the surface of the star. By solving the TOV equations for all possible values of $\varepsilon_c$, we obtain the mass $M$ and radius $R$ of the star. The mass function $M(r)=r(1-e^{-2\lambda(r)})/2$ satisfies Eq. (\ref{tov3}). Compactness ($C$) is a global parameter that quantifies the stellar gravity and it is defined in terms of the stellar mass $M$ and radius $R$ as $C=M/R$. 

We adopt the methodology to calculate the non-radial oscillations of the DMSQSs using the Cowling approximations, which is well presented in Ref.~\cite{Sotani:2010mx}. After solving the TOV Eqs. (\ref{tov}) - (\ref{tov3}), we obtain the oscillations mode frequencies by solving two following coupled differential equations:
\begin{eqnarray}
\frac{dW(r)}{dr}&=&\frac{d\varepsilon(r)}{dP(r)}\Bigg[\omega^2r^2e^{\lambda(r)-2\Phi(r)}V(r) + \frac{d\Phi(r)}{dr}W(r)\Bigg] - l(l+1)e^{\lambda(r)}V(r),
\label{W eqn}\\
\frac{dV(r)}{dr}&=&2\frac{d\Phi(r)}{dr}V(r) - e^{\lambda(r)}\frac{W(r)}{r^2}.
\label{V eqn}
\end{eqnarray}

In order to solve the Eqs. (\ref{W eqn}) and (\ref{V eqn}), the two boundary conditions at the center ($r=0$) and the surface ($r=R$) of the star are to be imposed. Near the center ($r=0$) of the star the functions $W(r)$ and $V(r)$ behave as
\begin{eqnarray}
W(r)=Ar^{l+1}~{\rm{and}}~~V(r)=-Ar^l/l;
\label{bc_center}
\end{eqnarray}
where $A$ is an arbitrary constant. Towards the surface ($r=R$) of the star the vanishing perturbation functions provide another boundary condition for $W(r)$ and $V(r)$ and is written as
\begin{eqnarray}
\omega^2R^2e^{\lambda(R)-2\Phi(R)}V(R) + \frac{d\Phi(r)}{dr}\Bigg\rvert_{r=R}W(R) = 0.
\label{bc_surface}
\end{eqnarray}
The coupled differential equations, Eqs.~(\ref{W eqn}) and (\ref{V eqn}), are integrated from the center to the surface of the star by assuming an initial value of $\omega^2$. After each integration the value of $\omega^2$ is improved using Ridders’ method until Eq.~(\ref{bc_surface}) is satisfied.
\par
The tidal deformability is an astrophysical observable that defines the quadrupole deformation of a compact star in a binary system due to the tidal field created by its companion star. In case of an isolated star, the tidal deformability is calculated in the limit where source of the static external quadrupolar tidal field is very far away \cite{Hinderer:2007mb}.
The dimensionless tidal deformability ($\Lambda$) is given as
\begin{eqnarray} 
\Lambda=\frac{2}{3} k_2 R^5.
\label{Lambda}
\end{eqnarray}
Here $k_2$ is the tidal love number, which is given in terms of the compactness ($C$), and a quantity ($y$) obtained by following the formalism developed in Ref.~\cite{Hinderer:2007mb,Hinderer:2009ca}. In case of QSs, $y$ is defined as \cite{Hinderer:2009ca,Kumar:2022byc}
\begin{eqnarray}
y=\frac{RH'(R)}{H(R)} - \frac{4\pi R^3 \varepsilon_s}{M(R)};
\label{y}
\end{eqnarray}
where, $\varepsilon_s$ is the energy density at the surface of the QS.
 

\section{Results}
\label{Results}

In this section, we discuss the results obtained for $f$-mode oscillations using the Cowling approximation (given in Sec. \ref{Structure}) for DMSQSs (given in Sec. \ref{Models}). First, we compute the DMSQSs EoS using Eqs. (\ref{e}) and (\ref{P}). With the DMSQSs EoS, we then obtain the static stellar structure by numerically solving the TOV equations given by Eqs. (\ref{tov}) to (\ref{tov3}) from the center to the surface of the star. We then numerically integrate the coupled differential Eqs. (\ref{W eqn}) and (\ref{V eqn}) from the center to the surface of the star using the appropriate boundary conditions given by Eqs. (\ref{bc_center}) and (\ref{bc_surface}). The $f$-mode frequencies are obtained with the required precision using Ridders’ method. We also calculate the tidal deformability $\Lambda$ of DM admixed QSs using the Eqs. (\ref{y}) and (\ref{Lambda}). In order to study the effect of DM for different values of the SQM vector coupling constant $G_V$, we vary the DM particle mass $m_\chi$ and the Fermi momentum of the DM fermions ${k_F}_{\chi}$ and study their effects on the stellar structure and $f$-mode oscillation of the DMSQSs. We have investigated the explicit dependence of the DMSQS properties on $m_{\chi}$ and ${k_F}_{\chi}$ by varying them individually. While varying $m_{\chi}$ and ${k_F}_{\chi}$ we have kept the vBag parameters like $G_V$ and corresponding value of $B_{av}$ fixed. For example, we fixed $G_V$=0.3 fm$^2$ and the corresponding value of $B_{av}$ when we varied $m_{\chi}$. Again we fix $G_V$=0.5 fm$^2$ with the corresponding value of $B_{av}$ while varying ${k_F}_{\chi}$. While varying $m_{\chi}$, we have fixed ${k_F}_{\chi}$ and vice versa. This way, the dependence of the properties of DMSQS on both the mass and the number density of DM are examined explicitly and individually in the present paper. While varying the parameters of dark sector, the parameters of the MIT vbag model are kept fixed in order to study the distinct role that DM plays on the properties of DMSQS.

\begin{figure*}[t]
\centering
\subfigure[\,Fixed ${k_F}_{\chi}$= 0.03 GeV and different $G_V$ and $m_{\chi}$.]{\includegraphics[width=0.48\linewidth,height=7cm]{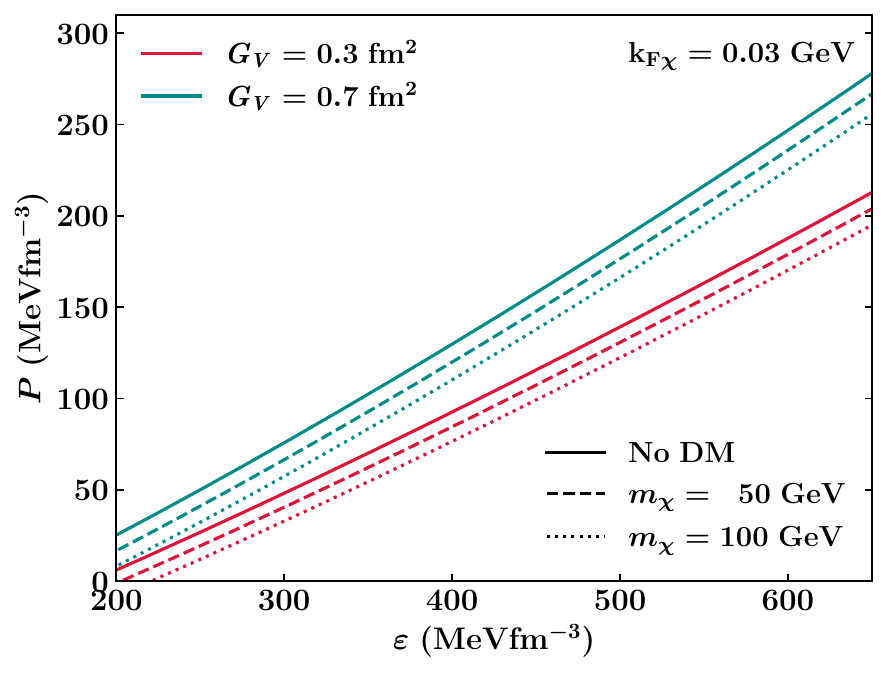} \label{fig:eosGVmchi}}\quad
\hfill 
\subfigure[\,Fixed $G_V$ = 0.5 fm$^2$ and $m_{\chi}$= 75 GeV and different ${k_F}_{\chi}$.]{\includegraphics[width=0.48\linewidth,height=7cm]{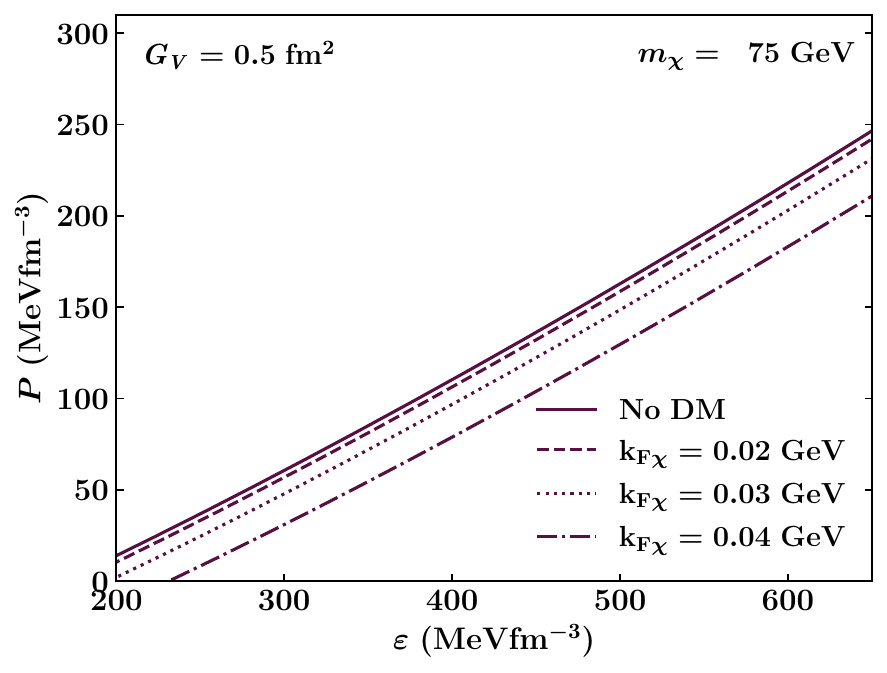}\label{fig:eoskfchi}}
\caption{Equation of state of dark matter admixed strange quark star.}
\label{fig:eos}
\end{figure*}

\begin{figure*}[ht]
\centering
\subfigure[\,Mass profiles by varying $G_V$, ${k_F}_{\chi}$, and $m_{\chi}$.]
{\includegraphics[width=0.48\linewidth,height=7cm]{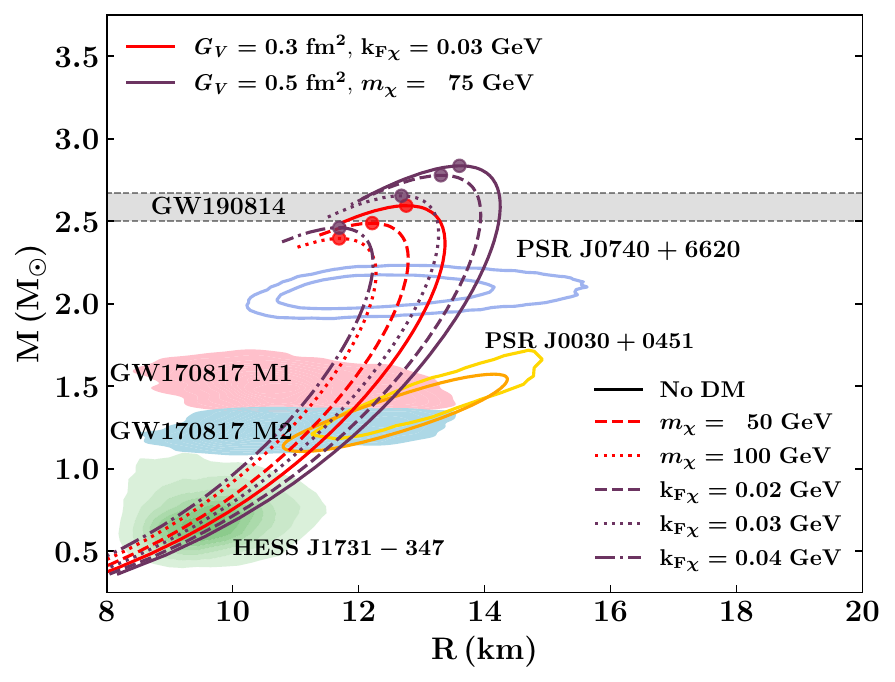} \label{fig:MRn}}\quad
\hfill 
\subfigure[\,Maximum mass region for different parameterizations by varying $G_V$, ${k_F}_{\chi}$, and $m_{\chi}$. We follow Fig.~\ref{fig:eos} for labels.]{\includegraphics[width=0.48\linewidth,height=7cm]{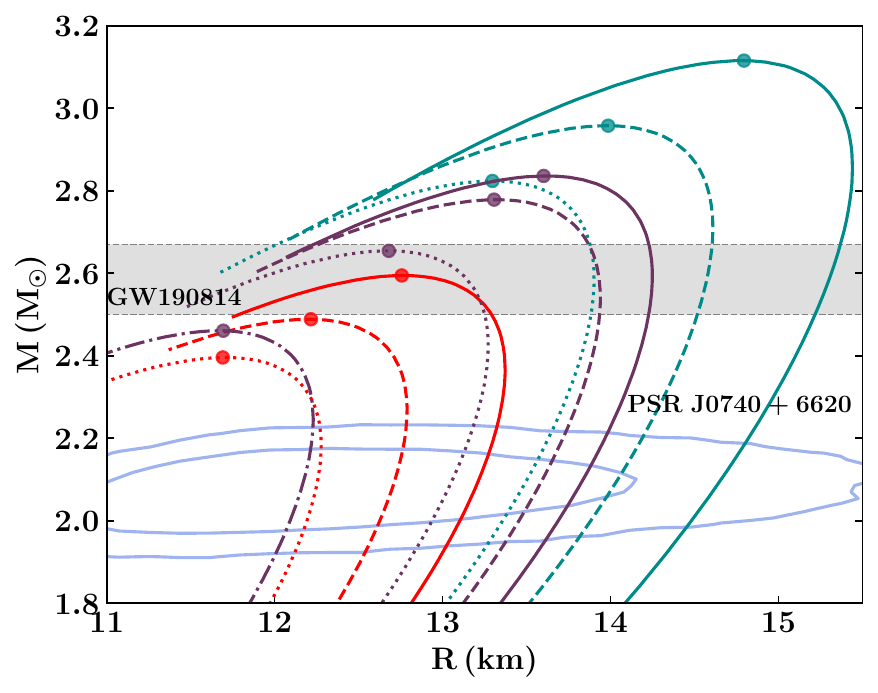}\label{fig:canMR}}
\caption{Variation of mass with radius of dark matter admixed strange quark stars (DMSQSs). The observational limits imposed on maximum mass from the most massive pulsar PSR J0740+6620 \cite{Fonseca:2021wxt} and corresponding radius \cite{Miller:2021qha, Riley:2021pdl} are also indicated. The constraints on $M-R$ plane prescribed from GW170817 \cite{LIGOScientific:2018cki}, the NICER experiment for PSR J0030+0451 \cite{Riley:2019yda, Miller:2019cac}, and HESS J1731-347 \cite{Doroshenko2022} are also compared. The mass of the secondary component of GW190814 \cite{LIGOScientific:2020zkf} is also displayed.}
\label{fig:MR}
\end{figure*}
\begin{figure*}[ht]
\centering
\subfigure[\,Fixed ${k_F}_{\chi}$= 0.03 GeV and $G_V$= 0.3 fm$^2$ and different $m_{\chi}$.]{\includegraphics[width=0.48\linewidth,height=7cm]{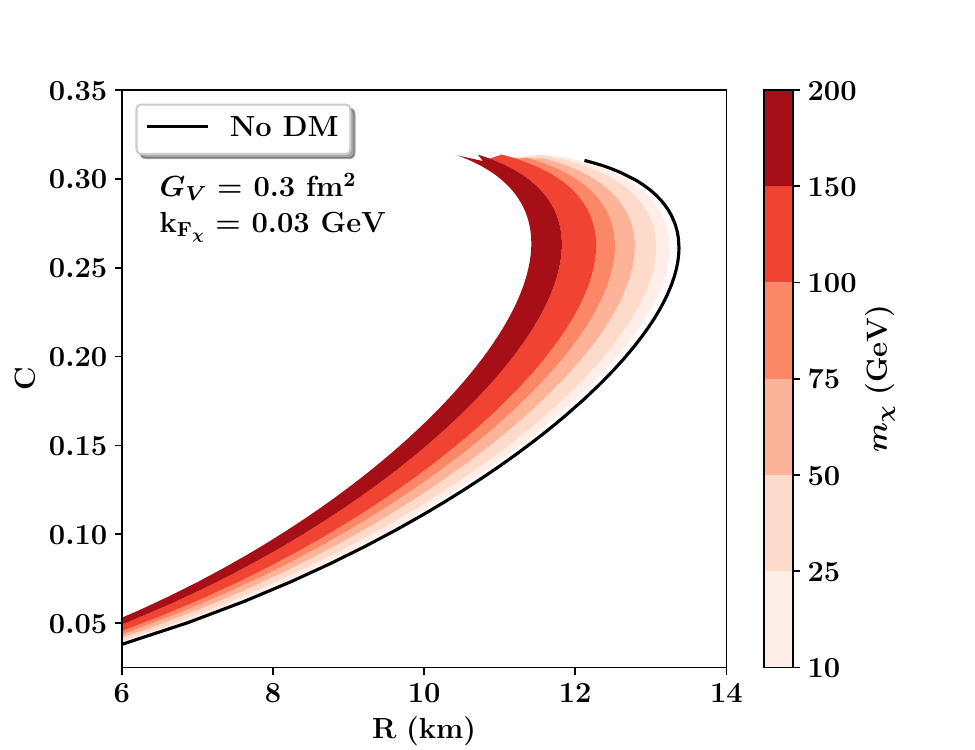} \label{fig:CRGVmchi}}\quad
\hfill 
\subfigure[\,Fixed $G_V$= 0.5 fm$^2$ and $m_{\chi}$= 75 GeV and different ${k_F}_{\chi}.$]{\includegraphics[width=0.48\linewidth,height=7cm]{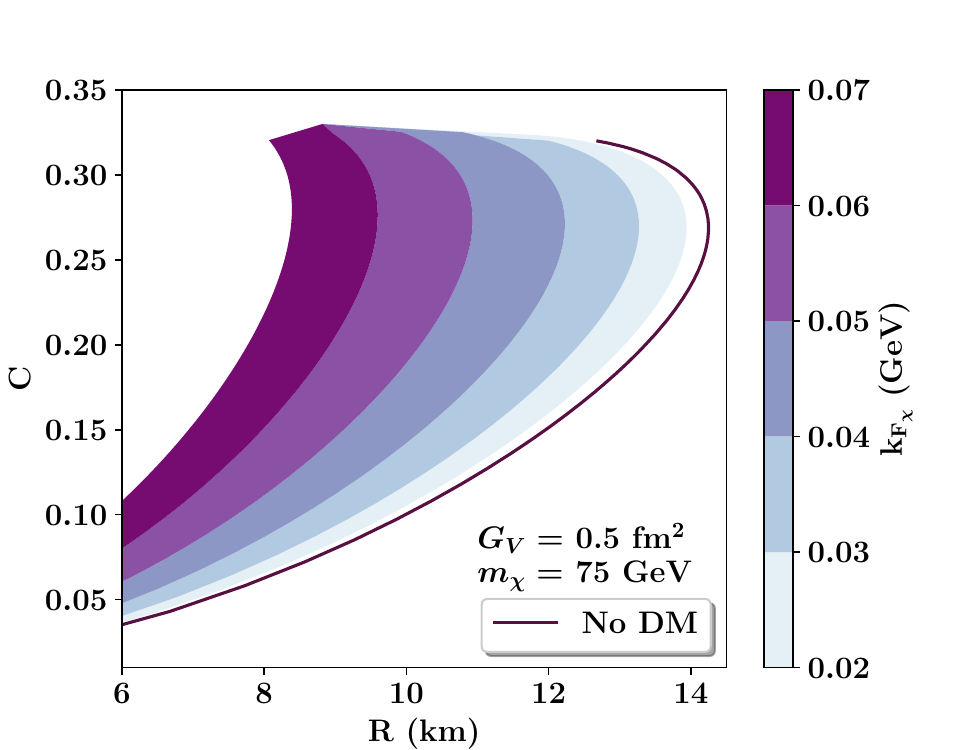}\label{fig:CRkfchi}}
\caption{Variation of compactness $C=M/R$ with radius of DMSQSs.}
\label{fig:CR}
\end{figure*}

We first study the effects of variation of $m_{\chi}$ on the EoS and structural properties of the DMSQSs for a fixed value of ${k_F}_{\chi}$=0.03 GeV and by considering different values of $G_V$. Later, we will investigate the effects of variation of ${k_F}_{\chi}$ for fixed $G_V$ and $m_{\chi}$ values. The EoS of the DMSQSs for different values of $m_{\chi}$= 50 and 100 GeV and $G_V$= 0.3 and 0.7 fm$^2$ are shown in Fig.~\ref{fig:eosGVmchi}. 
We can see that the higher values of $G_V$ lead to stiffer EoS and consequently massive SQSs with larger radius (see also Fig. \ref{fig:MR}). This trend is similar to that seen in Ref.~\cite{Lopes:2020btp}. So, for any particular value of $m_{\chi}$, stronger coupling strength between quarks gives massive DMSQSs with larger radius. Also, for any fixed value of $G_V$, Fig. \ref{fig:eosGVmchi} shows that the EoS softens with higher values of $m_{\chi}$. This is also reflected in the structural properties of the DMSQSs as seen in Fig. \ref{fig:MR}.  We can see that the mass and radius of the DMSQSs reduce for massive DM. The reason can be attributed to the fact that presence of DM makes the star more compact and the heavier fermionic DM makes the star even more compact. To illustrate this, we show the variation of dimensionless compactness ($C$) with respect to radius in Fig. \ref{fig:CRGVmchi} for $G_V$=0.3 fm$^2$. It can be inferred that, for a particular value of radius, $C$ increases with $m_{\chi}$ in order to incorporate massive DM at constant ${k_F}_{\chi}$ or constant $\rho_{\chi}$. Thus, lighter DM yields massive but less compact DMSQSs. For a low value of $m_{\chi}$, not only the mass but also the radius of the DMSQS increase which makes the star less compact. For example, the combination $G_V$=0.3 fm$^2$ and $m_{\chi}$=50 GeV can produce a 1.4 $M_{\odot}$ DMSQS of corresponding radius $R_{1.4}$=11.60 km and corresponding compactness $C$=0.178; while $m_{\chi}$=100 GeV produces a 1.4 $M_{\odot}$ DMSQS of radius $R$=11.30 km and corresponding compactness $C$=0.183. 

Next, in Fig. \ref{fig:eoskfchi}, we present the EoS of DMSQSs for different values of ${k_F}_{\chi}$ for fixed $G_V$ = 0.5 fm$^2$ and $m_{\chi}$= 75 GeV. We notice that, with $G_V$=0.5 fm$^2$, $m_{\chi}$=75 GeV, the EoS softens with increasing values of ${k_F}_{\chi}$ or $\rho_{\chi}$. This is because if the QS is largely populated by fermionic DM, the compactness $C$ increases as seen from Fig. \ref{fig:CRkfchi}, leading to softening of the EoS and consequently decreasing the mass and radius of the star, as seen from Fig. \ref{fig:MR}. For example, with fixed $G_V$=0.5 fm$^2$ and $m_{\chi}$=75 GeV we obtain a 1.4 $M_{\odot}$ DMSQS of radius $R_{1.4}$=11.12 km and corresponding compactness $C$=0.186 for ${k_F}_{\chi}$=0.04 GeV but ${k_F}_{\chi}$=0.02 GeV produces a 1.4 $M_{\odot}$ DMSQS of radius $R$=12.24 km and corresponding compactness $C$=0.169. So, SQSs that are largely populated with massive DM fermions, are highly compact with comparatively less masses and radii.

We also find from Fig. \ref{fig:MRn} that the various astrophysical constraints on the $M-R$ relation of compact stars from the most massive PSR J0740+6620, GW170817, the NICER data for PSR J0030+0451, and HESS J1731-347 are well satisfied with the DMSQS EoS. We also observe that the mass of the secondary component of GW190814 \cite{LIGOScientific:2020zkf} is also satisfied by the SQSs and with $G_V>$ 0.3 fm$^2$ for the DMSQS. However, the exact nature of the massive secondary component of GW190814 is not known; whether it is a NS/QS or a black hole. As also seen in Ref.~\cite{Sen:2022pfr}, all these constraints are better satisfied for higher values of $G_V$, since in Fig. \ref{fig:MRn}, we find that for $G_V$=0.3 fm$^2$ the NICER data for PSR J0030+0451 are violated for $m_{\chi}$=100 GeV. However, in Ref.~\cite{Sen:2022pfr}, the recently obtained constraint from HESS J1731-347 \cite{Doroshenko2022} was not included. Therefore, in the present work we also test the DMSQSs EoS in the light of this constraint obtained from the low mass pulsar with low radius. We find that this constraint is successfully satisfied with our compact DMSQSs configurations. On the other hand, for $G_V$=0.5 fm$^2$, $m_{\chi}$=75 GeV, shows that low DM content with low ${k_F}_{\chi}$ or $\rho_{\chi}$ can better satisfy the observational constraints since ${k_F}_{\chi}$= 0.04 GeV makes the star too compact to satisfy the NICER data for PSR J0030+0451. We also show the maximum mass region for various parameterisations of DMSQS in Fig.~\ref{fig:canMR}. Here, we have included all the configurations discussed in this work and it follows the label given in Fig. \ref{fig:eos}. 
As discussed earlier, an increase in the value of $G_V$ results in a stiffer EoS and this results in a larger maximum mass value. Similarily, we also find that with an increase in ${k_F}_{\chi}$ and $m_{\chi}$ values, the EoS becomes less stiff and results in stellar profiles with lower maximum mass values. 
\par 
\begin{figure*}[ht]
\centering
\subfigure[\,Fixed ${k_F}_{\chi}$= 0.03 GeV and different $G_V$ and $m_{\chi}$.]{\includegraphics[width=0.48\linewidth,height=7cm]{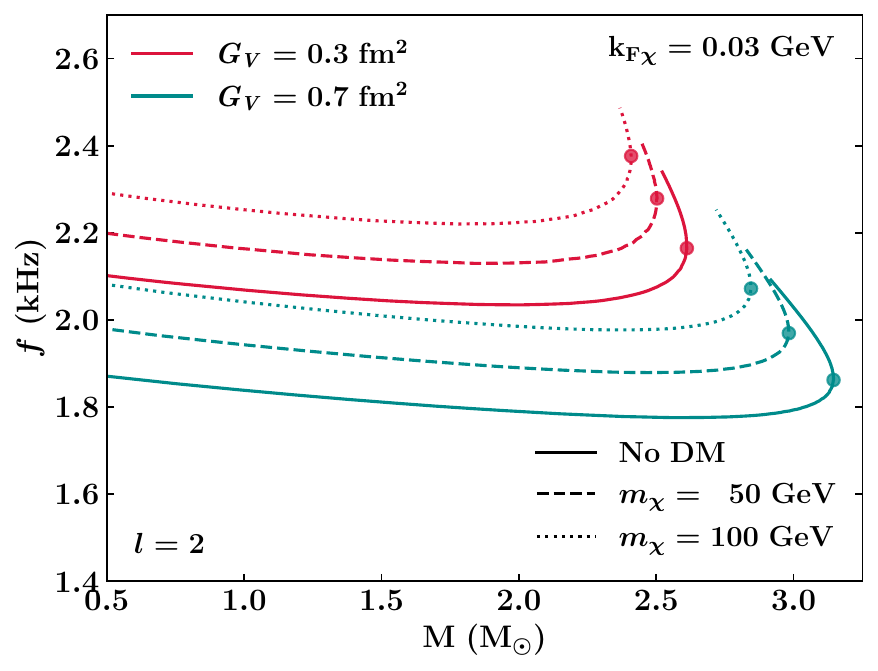} \label{fig:fMGVmchi}}\quad
\hfill 
\subfigure[\,Fixed $G_V$= 0.5 fm$^2$ and $m_{\chi}$= 75 GeV and different ${k_F}_{\chi}$.]{\includegraphics[width=0.48\linewidth,height=7cm]{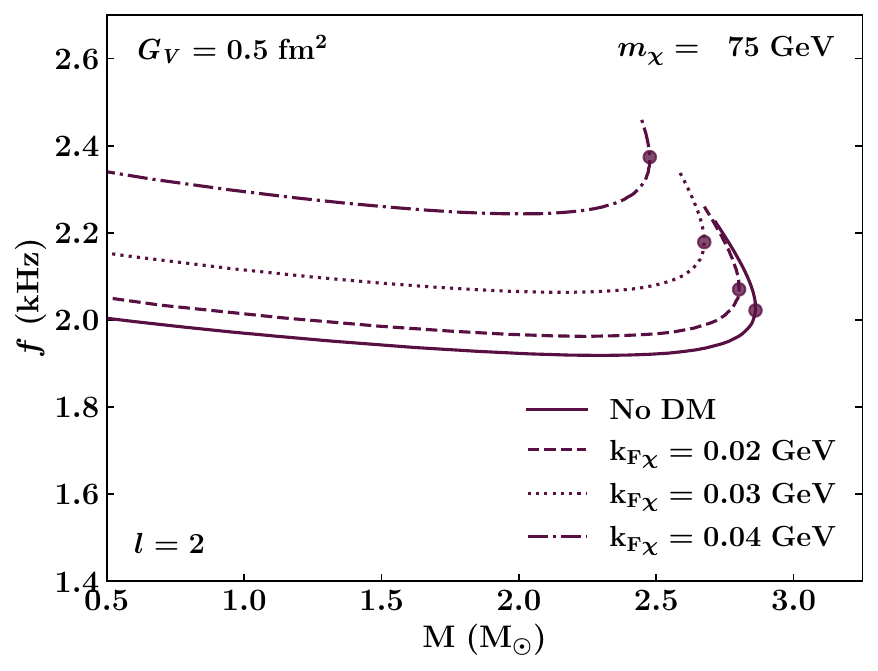}\label{fig:fMkfchi}}
\caption{Variation of $f$-mode frequency with mass of dark matter admixed strange quark star (DMSQS).}
\label{fig:fM}
\end{figure*}
\begin{figure*}[ht]
\centering
\subfigure[\,Fixed ${k_F}_{\chi}$= 0.03 GeV and different $G_V$ and $m_{\chi}$.]{\includegraphics[width=0.48\linewidth,height=7cm]{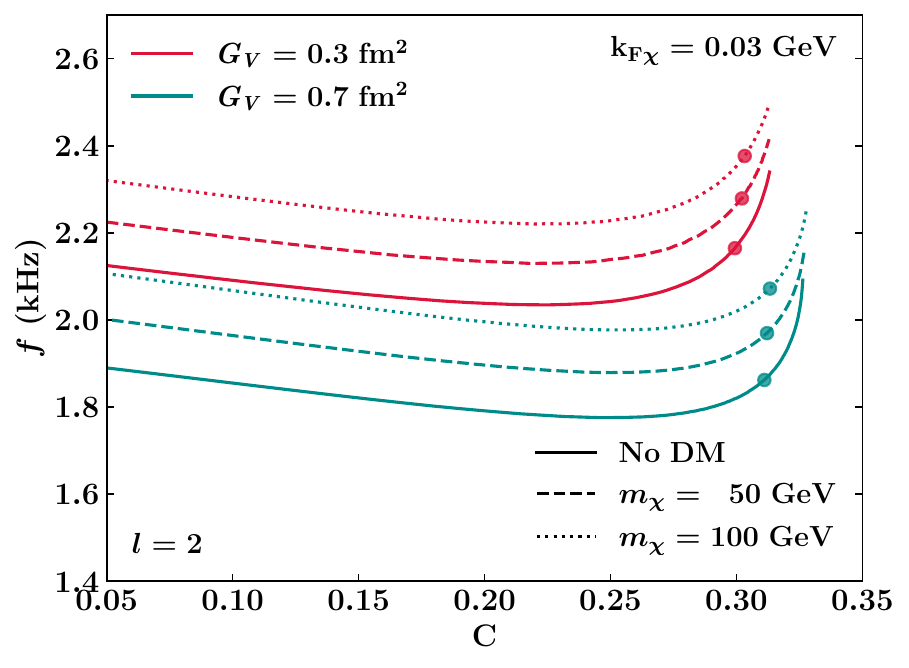} \label{fig:fCGVmchi}}\quad
\hfill 
\subfigure[\,Fixed $G_V$= 0.5 fm$^2$ and $m_{\chi}$= 75 GeV and different ${k_F}_{\chi}$.]{\includegraphics[width=0.48\linewidth,height=7cm]{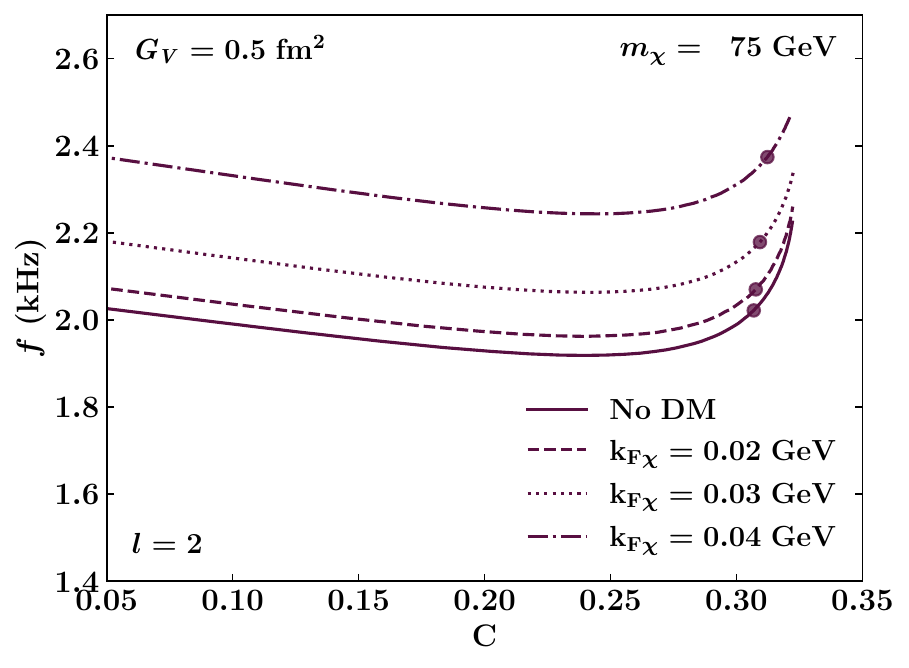}\label{fig:fCkfchi}}
\caption{Variation of $f$-mode frequency with compactness of DMSQS.}
\label{fig:fC}
\end{figure*}
We now extend our work to focus on the $l=2$ $f$-mode oscillations of DMSQSs in Fig.~\ref{fig:fM}, where we plot the $f$-mode frequencies as a function of stellar mass by fixing the value of the coupling constant $G_V$=0.3, 0.7 fm$^2$ and the Fermi momentum of DM fermions ${k_F}_{\chi}=0.03$ GeV (Fig. \ref{fig:fMGVmchi}) for $m_{\chi}$= 50 and 100 GeV. We find that, the $f$-mode frequencies of DMSQSs decrease slightly with the mass and it only starts to increase towards the maximum mass of the star. With increase in value of $G_V$, we observe an overall decrement in the $f$-mode spectra of DMSQS. For any value of $G_V$, the $f$-mode frequencies show an overall increment with increase in the value of $m_{\chi}$. In the `no-DM' (pure SQS) scenario, we find that larger vector repulsion between the quarks leads to lowering of the $f$-mode frequency of QSs.
For example, a 1.4 $M_{\odot}$ SQS with $G_V$= 0.3 fm$^2$ has $f$-mode oscillation frequency as high as 2.05 kHz compared to that with $G_V$= 0.7 fm$^2$ having $f$= 1.81 kHz. In the presence of DM, lighter fermionic DM yields more massive DMSQSs with lower value of $f$-mode oscillation frequency. For example, with $G_V$=0.7 fm$^2$, the 1.4 $M_{\odot}$ DMSQS produced for $m_{\chi}$=50 GeV oscillates with frequency $f$=1.92 kHz but $m_{\chi}$=100 GeV yields a 1.4 $M_{\odot}$ DMSQS with oscillation frequency $f$=2.02 kHz.
Next, in Fig. \ref{fig:fMkfchi}, we plot the $f$-mode oscillations of DMSQSs by fixing $G_V=0.5$ fm$^2$ and $m_\chi = 75$ GeV, for different values of ${k_F}_{\chi}=0.02, 0.03$ and 0.04 GeV. We observe that the $f$-mode spectra are more sensitive to the variations in ${k_F}_{\chi}$, compared to the changes in $m_\chi$ values. It can be seen that, larger values of ${k_F}_{\chi}$ results in overall increment in $f$-mode oscillations, compared to the no-DM case. For a 1.4 $M_{\odot}$ DMSQS, with ${k_F}_{\chi}$= 0.02 (0.04) GeV, we have $f_{1.4}$= 1.99 (2.09) kHz.
\par 
We now study the explicit dependence of $f$-mode frequency of DMSQSs on the compactness of the star in Fig.~\ref{fig:fC}. As we have already seen that the presence of DM largely affects the compactness of the SQSs (Fig. \ref{fig:CR}), it is expected that this will be reflected in $f$-modes. For any fixed value of $C$, the $f$-mode frequency is found to be higher for lower coupling constant $G_V$ and for higher mass values of fermionic DM $\chi$ (see Fig.~\ref{fig:fCGVmchi}). Similar to $m_\chi$, an increment in ${k_F}_{\chi}$ results in a more sensitive enhancement of $f$-modes as shown in Fig.~\ref{fig:fCkfchi}. Here, we conclude that, highly compact DMSQSs, having massive DM (Fig. \ref{fig:fCGVmchi}) or high DM content in terms of high ${k_F}_{\chi}$ (Fig. \ref{fig:fCkfchi}), oscillate faster with higher values of $f$-mode oscillation frequency.
\par
\begin{figure*}[ht]
\centering
\subfigure[\,Tidal deformability values of DMSQS by varying $G_V$, ${k_F}_{\chi}$, and $m_{\chi}$.]{\includegraphics[width=0.48\linewidth,height=7cm]{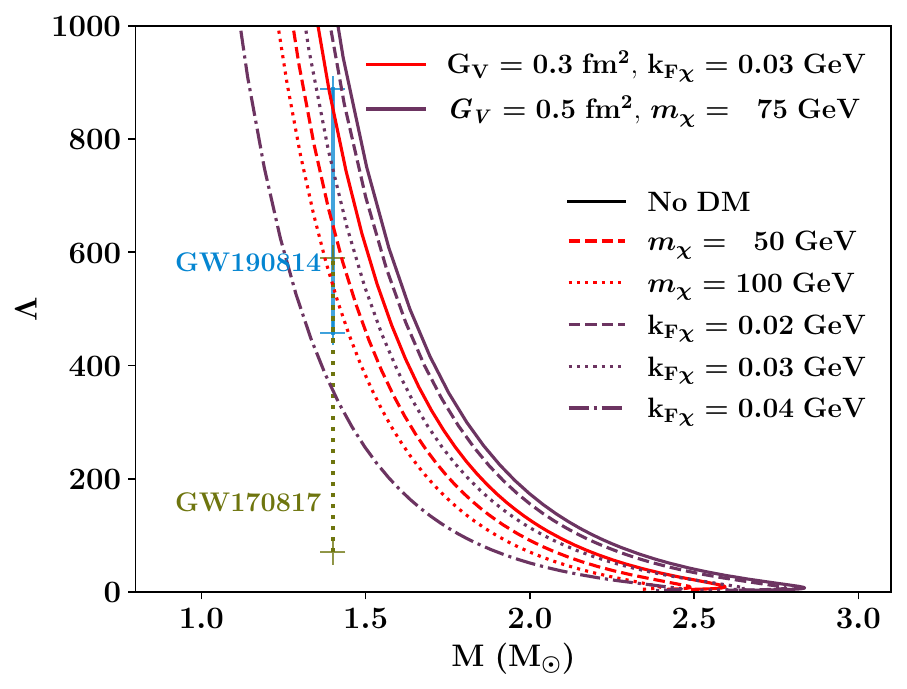} \label{fig:LM}}\quad
\hfill 
\subfigure[\,Tidal deformability values of DMSQS near to $1.4M_{\odot}$ profile. For labels refer Fig.~\ref{fig:eos}.]
{\includegraphics[width=0.48\linewidth,height=7cm]{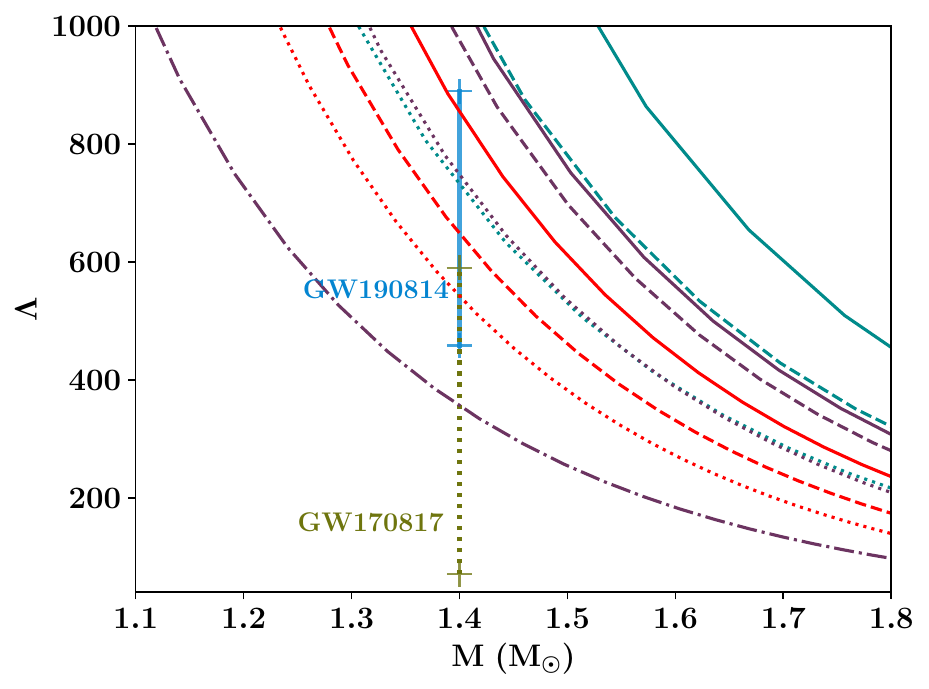}\label{fig:canLM}}
\caption{ Variation of tidal deformability with mass of DMSQS. The constraints on $\Lambda_{1.4}$ from GW170817 \cite{LIGOScientific:2018cki} and GW190814 \cite{LIGOScientific:2020zkf} are also shown.}
\label{fig:Lambda}
\end{figure*}
\begin{figure*}[ht]
\centering
\subfigure[\,Fixed ${k_F}_{\chi}$= 0.03 GeV and different $G_V$ and $m_{\chi}$.]{\includegraphics[width=0.48\linewidth,height=7cm]{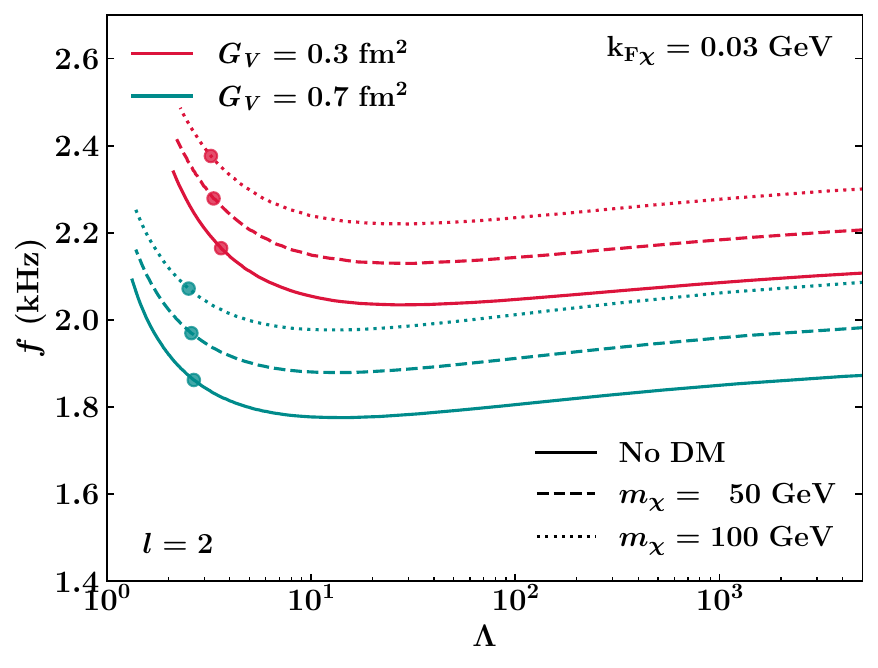} \label{fig:fLGVmchi}}\quad
\hfill 
\subfigure[\,Fixed $G_V$= 0.5 fm$^2$ and $m_{\chi}$= 75 GeV and different ${k_F}_{\chi}$.]{\includegraphics[width=0.48\linewidth,height=7cm]{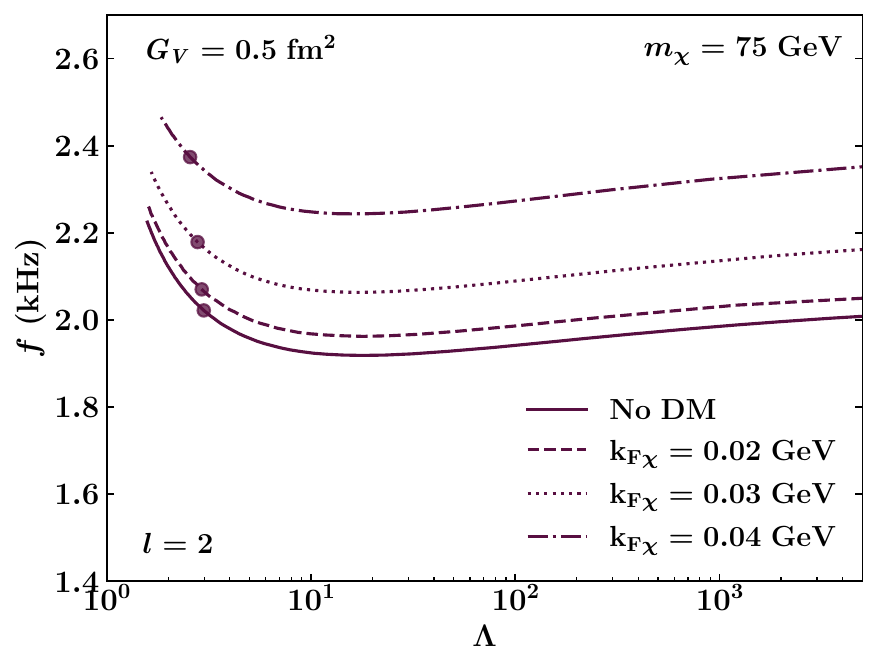}\label{fig:fLkfchi}}
\caption{ Variation of $f$-mode frequency with tidal deformability of dark matter admixed strange quark star (DMSQS).}
\label{fig:fL}
\end{figure*}
%
We plot the tidal deformability $\Lambda$ as a function of stellar mass by varying $G_V$, $m_{\chi}$, and ${k_F}_{\chi}$ values in Fig.~\ref{fig:LM}. We find that the value of $\Lambda$ decreases with an increase in stellar mass till it attains the maximum mass. Also, in Fig.~\ref{fig:canLM}, we show the tidal deformability values close to $1.4M_\odot$ ($\Lambda_{1.4}$) star along with the constraints given by GW170817 and GW190814. We have included all the parameterizations of DMSQSs discussed in this work and it follows the labels given in Fig.~\ref{fig:eos}. We find that the constraint on $\Lambda_{1.4}$ from GW170817 is satisfied with $m_{\chi}$=100 GeV and $G_V$=0.3 fm$^2$. For the no-DM case as well as the DMSQSs with $m_{\chi}$=50 GeV, this constraint is not at all satisfied for any values of $G_V$ because we obtain very massive DMSQS configurations whose tidal deformability is also very high, thereby violating this constraint. However, the constraint from GW190814 on $\Lambda_{1.4}$ is comparatively better satisfied for the parameters considered for DMSQSs in this work. In Fig. \ref{fig:LM}, we find that this constraint is well satisfied by both SQS and DMSQSs for $G_V$=0.3 fm$^2$. However, if there is strong repulsion among the quarks via high vector quark coupling strength then we require the presence of DM, rather massive DM to satisfy this constraint. The combined effect of the strong interaction between the quarks balanced by the feeble SQM-DM interaction helps to reduce the mass, radius and hence the tidal deformability of the DMSQSs, thereby helping to satisfy the constraint from GW190814 on $\Lambda_{1.4}$. We also study the relation between tidal deformability and mass of the SQSs for the variation of ${k_F}_{\chi}$ for fixed $G_V$ and $m_{\chi}$. We notice that in order to satisfy the constraints on $\Lambda_{1.4}$ from GW190814, high content of DM is necessary by the SQSs. Moderately massive ($m_{\chi}$ = 75 GeV) DM of Fermi momentum as high as ${k_F}_{\chi}$= 0.04 GeV is also capable of satisfying the constraint from GW170817 on $\Lambda_{1.4}$. However, we have seen from Fig. \ref{fig:MRn} that this combination of ($G_V$, $m_{\chi}$, ${k_F}_{\chi}$)=(0.5 fm$^2$, 75 GeV, 0.04 GeV) do not satisfy the NICER data for PSR J0030+0451. So an intermediate value of ${k_F}_{\chi}$ between 0.03 and 0.04 GeV may satisfy both the constraints from the NICER data for PSR J0030+0451 as well as that on $\Lambda_{1.4}$ from GW170817. 
\par
%
%
\begin{figure*}
\centering
\subfigure[\,
Linear relation for $f$-mode oscillations and average density.
]{\includegraphics[width=0.48\linewidth,height=7cm]{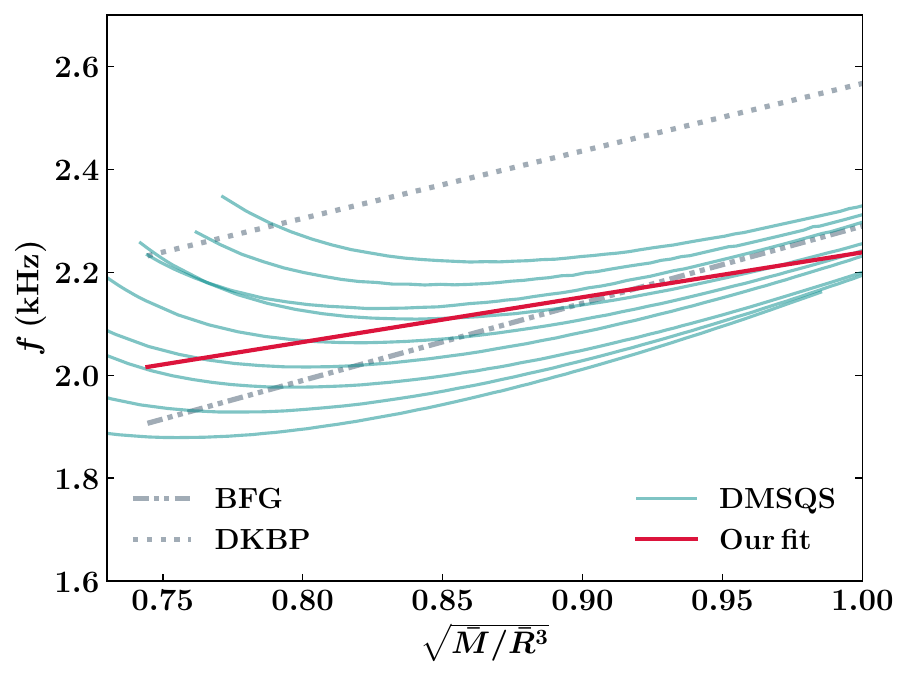} \label{fig:fit-plot}}\quad
\hfill 
\subfigure[\,
Angular frequency ($\omega=2\pi f$) normalized by stellar mass ($M$) as a function of compactness $C$.
]{\includegraphics[width=0.48\linewidth,height=7cm]{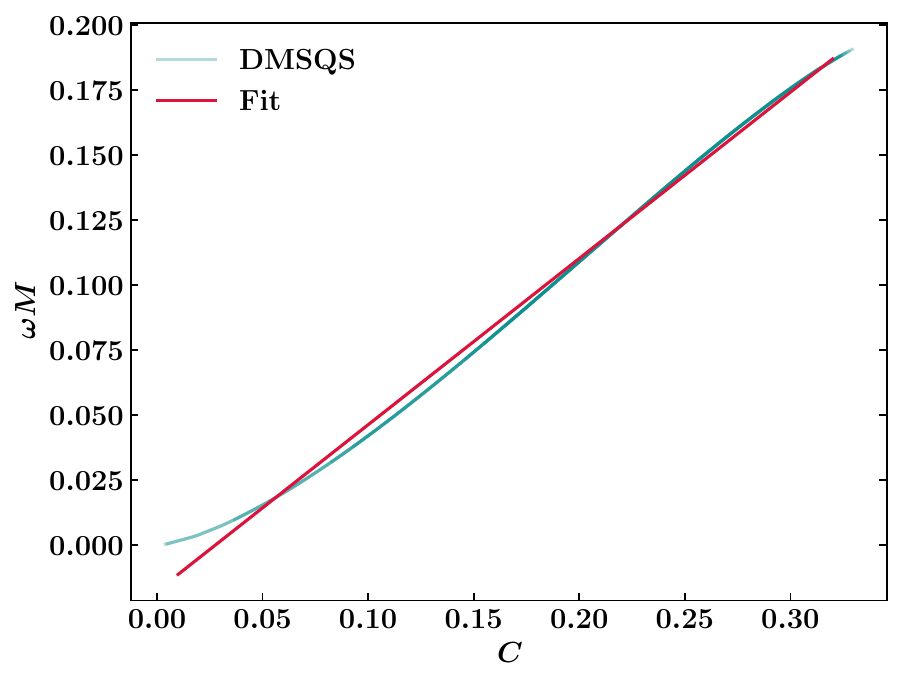}\label{fig:univ-rel}}
\caption{(a) Variation of $f$-mode frequency with average density $\bar{\rho}=\sqrt{\bar{M}/\bar{R}^3}$ of DMSQSs. The solid red line shows the linear fit obtained for different parameterizations of DMSQSs used in this work. We also plot the linear fit obtained by Das et al.~\cite{Das:2021dru} denoted by (DKBP) and that obtained in Benhar et al. ~\cite{Benhar:2004xg} represented with (BFG), for comparison. (b) Variation of angular frequency $\omega$ normalised by stellar mass $M$ with compactness $C$. The solid red line represents the universal relation obtained for DMSQSs.}
\label{fig:lin-fit}
\end{figure*}
Further, in Fig. \ref{fig:fL}, we analyse the variation of the $f$-mode frequency of the DMSQSs with respect to its tidal deformability ($\Lambda$). We observe that, the mode frequencies slightly decrease with decrement in $\Lambda$ and it only begins to increase towards the $\Lambda$ corresponding to maximum mass of the star. We plot the $f$-modes for fixed ${k_F}_{\chi}$ and different $G_V$, $m_{\chi}$ in Fig. \ref{fig:fLGVmchi}. As shown in Figs.~\ref{fig:fMGVmchi} and \ref{fig:fCGVmchi}, the $f$-mode frequencies decrease for high values of $G_V$; while for a fixed $G_V$, the frequencies increase with increment in $m_\chi$. In Fig. \ref{fig:fLkfchi}, we plot the $f$-modes as a function of $\Lambda$ for fixed $G_V$,  
$m_{\chi}$ and by varying ${k_F}_{\chi}$. As observed in Figs.~\ref{fig:fMkfchi} and \ref{fig:fCkfchi}, $f$-mode oscillations are sensitive to variation in ${k_F}_{\chi}$ and large values of ${k_F}_{\chi}$ results in higher values of frequencies.

\par
Finally, for different parameterizations of DMSQSs, we obtain a linear (empirical) relation between the $f$-mode frequency and the average density $\bar{\rho}=\sqrt{\bar{M}/\bar{R}^3}$, where $\bar{M}=M/(1.4$ $M_\odot$) and $\bar{R}=R/(10$ km). The empirical formula of the form, $f=a+b\bar{\rho}$ was first obtained by Andersson and Kokkotas~\cite{Andersson:1996pn} and by knowing the values of $f$-mode frequencies, we can calculate the mass and radius of the star. This could therefore help in constraining the EoS more closely~\cite{Andersson:1996pn}. For our EoSs, we obtain the values of $a$ and $b$ as 1.367 kHz and 0.87 kHz respectively. In Fig. \ref{fig:fit-plot}, we plot the relation obtained for DMSQSs along with the empirical relations from relevant earlier works for stellar models with dark matter admixed hyperon stars (DKBP)~\cite{Das:2021dru} and hybrid stars (BFG)~\cite{Benhar:2004xg}. We find that our fit relation differ considerably from the already existing relations from DKBP and BFG. However, our fit is found to be more close to the fit for hybrid 
stars denoted by BFG. 
Additionally, there are other model independent relations that provide a stronger correlation between the $f$-mode frequency, and both compactness and tidal deformability. First, we obtain the correlation between the angular frequency scaled by mass ($\omega M$) and the compactness ($C$) of the DMSQSs. In Ref.~\cite{Wen:2019ouw}, Wen et al. found that the variation of $\omega M$ as a function of $C$ is universal for 
$f$-modes. Similarly, in Fig.~\ref{fig:univ-rel}, we fit the mass scaled $\omega$ with $C$ for different parameterizations of DMSQSs. The universal relation was found to take the form $\omega M$ $= 0.634\,(M/R) - 0.0176$, where the coefficients are obtained after performing a linear fit. In order to obtain these correlations we varied the values of $G_V$ from 0.3 to 0.7 GeV and $m_\chi$ from 50 to 100 GeV.

\begin{figure}[h!]
    \centering
    \includegraphics[width=0.48\linewidth,height=7cm]{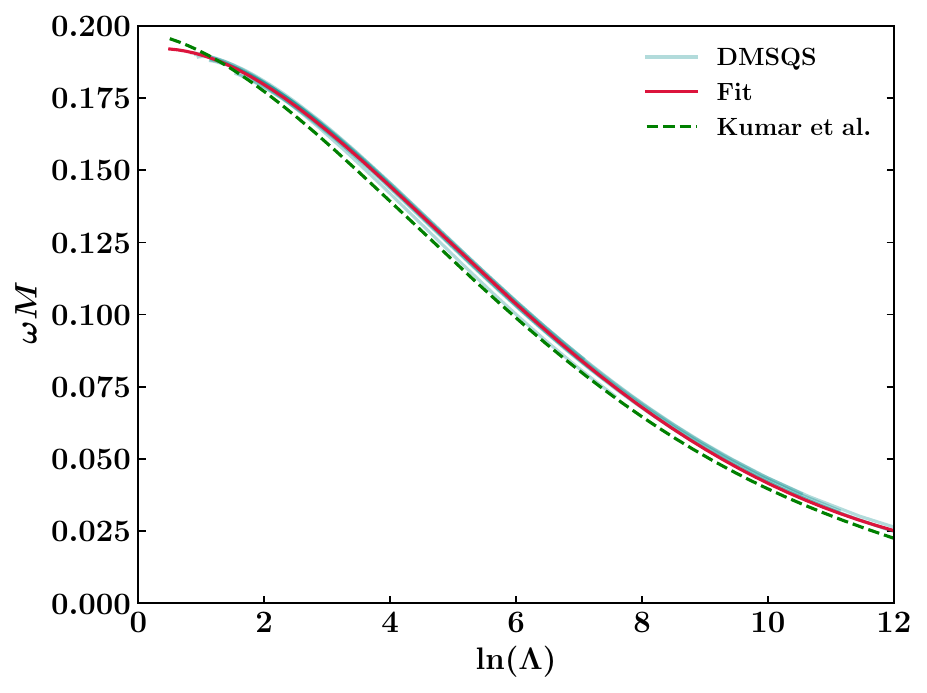}
    \caption{Mass scaled angular velocity ($\omega M$) as a function of tidal deformability ($\ln{\Lambda}$) of DMSQSs. The solid red line represents the correlation obtained by including all parameterizations of DMSQSs used in the paper. A correlation between $f$-mode frequency and $\Lambda$ obtained by Kumar et al.~\cite{Kumar:2023ojk} is also shown.}
    \label{fig:uni-L}
\end{figure}
\par
Further, one can also obtain a relation between the mass-scaled angular velocity and tidal deformability, where $\omega M$ is found to vary universally with $\Lambda$ and the relation is given by~\cite{Chan:2014kua,Sotani:2021kiw,Pradhan:2022vdf,Shirke:2024ymc}
\begin{equation}
    \omega M = \sum\limits_{i} \alpha_{i} (\ln{\Lambda})^{i}.
\end{equation}
In Fig.~\ref{fig:uni-L}, we plot $\omega M$ as a function of $\Lambda$ for different parameterization of DMSQSs. The correlation obtained is $0.191 + 0.004(\ln{\Lambda}) - 6.15\times 10^{-3}(\ln{\Lambda})^2 + 6.56\times 10^{-4}(\ln{\Lambda})^3  - 2.694\times 10^{-5}(\ln{\Lambda})^4 + 3.905\times 10^{-7}(\ln{\Lambda})^5$. We have also plotted the relation obtained between $\omega M$ and $\ln{\Lambda}$ given in Ref.~\cite{Kumar:2023ojk}, using the Cowling approximation: $0.198 - 2.612\times10^{-3}(\ln{\Lambda}) - 4.899\times 10^{-3}(\ln{\Lambda})^2 + 5.547\times 10^{-4}(\ln{\Lambda})^3  - 2.246\times 10^{-5}(\ln{\Lambda})^4 + 2.754\times 10^{-7}(\ln{\Lambda})^5$.

We find that both results are in good agreement. Some of the previous works on stars containing quark matter have also obtained the relation between $\omega M$ and $\Lambda$: $\omega M = 0.182 -6.11 \times 10^{-3}(\ln{\Lambda})- 4.594\times 10^{-3}(\ln{\Lambda})^2 +6.066\times 10^{-4}(\ln{\Lambda})^3 - 2.614\times 10^{-5}(\ln{\Lambda})^4 +2.228\times 10^{-7}(\ln{\Lambda})^5$~\cite{Pradhan:2023zmg} and $\log(2\pi f/\Lambda) = 1.052 - 0.984\,\log(M_{1.4}/\Lambda) - 0.030\,(\log(M_{1.4}/\Lambda))^2$~\cite{Ranea-Sandoval:2022izm}. 
\begin{figure}[h!]
    \centering
    \includegraphics[width=0.48\linewidth,height = 7cm]{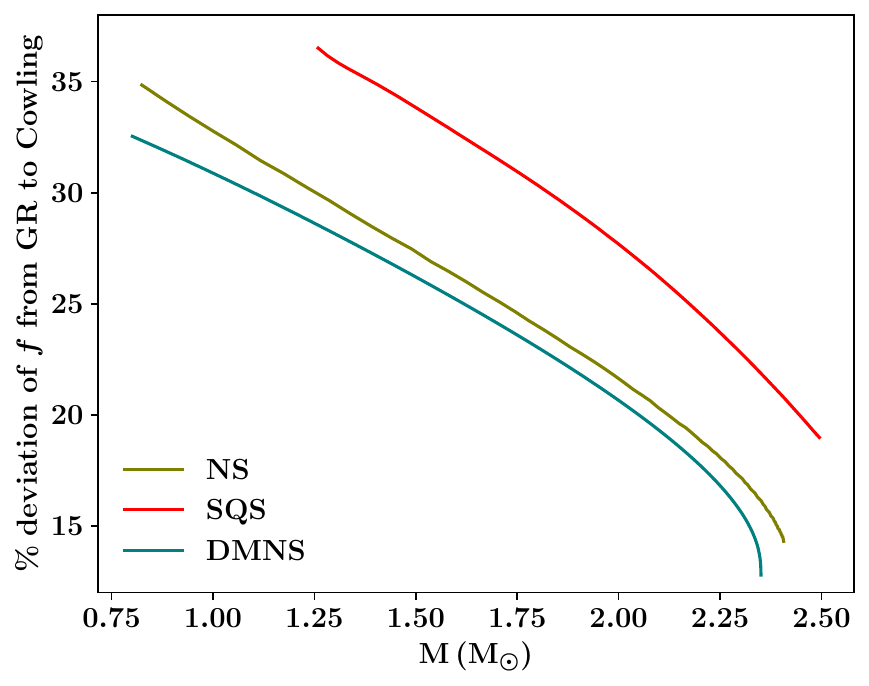}
    \caption{Percentage deviation of $f$-mode frequencies obtained using the Cowling approximation from the fully relativistic treatment for SQS with coupling constant $G_V=0.3$ fm$^{2}$~\cite{Lopes:2024yuu} and dark matter admixed neutron star (DMNS) with self-interaction strength $G=100$ fm$^{2}$~\cite{Shirke:2024ymc}.}
    \label{fig:dev-cow}
\end{figure}

Finally, we would like to comment on the usage of Cowling approximation in the case of strange quark stars (SQSs) by comparing with the results of fully relativistic analysis. 
Recently, the $f$-mode oscillation of SQSs modeled using vMIT Bag model with the same parameter set used in our analysis has been studied within the fully relativistic formalism~\cite{Lopes:2024yuu}.
We use their result to calculate the percentage deviation of $f$-mode frequencies obtained within the Cowling approximation from the fully general relativistic treatment ($\%$ deviation $= 100\times \lvert f_{Cow}-f_{GR}\rvert/f_{GR}$) for SQS with $G_V=0.3$ fm$^{2}$. The calculated deviation is shown as a function of stellar mass in Fig. \ref{fig:dev-cow}. We find that for massive SQS, the deviation percentage is closer to $20\%$, while for less massive stars, the deviation increases to a value of $30\%$. 
We also note that the deviation percentage decreases for SQS with high compactness. Further, as seen from Fig.~\ref{fig:CR}, the compactness increases with the introduction of DM. Thus, we believe that the inclusion of DM may not result in a higher deviation percentage for DMSQSs. 
Additionally, we also plot the percentage deviation using the result obtained for $f$-mode oscillations for neutron star (NS) and DM admixed neutron star (DMNS) with self interaction strength $G = 100$ fm$^{2}$, using the fully relativistic formalism from Ref.~\cite{Shirke:2024ymc}.  Our analysis shows that for the DMNS, the deviation percentage is smaller compared to the NS, indicating that the inclusion of dark matter may lead to a lower percentage of deviation. Thus, we expect that the inclusion of DM may decrease the deviation percentage of DMSQSs as well. Therefore, with the help of future gravitational wave detectors, we may be able to obtain more insights about the stellar properties of DMSQSs.


\section{Summary and Conclusion}
\label{Conclusion}

In the present work, we have investigated the non-radial $l=2$ $f$-mode oscillations of strange quark stars (SQSs) admixed with dark matter (DM). For our analysis, we have considered the presence of DM in SQSs, with the interaction between the fermionic DM $\chi$ and SQM $\psi_f$ being mediated by the dark mediator $\xi$. We note that, to the best of our knowledge, this is the first time the non-radial $f$-mode spectra of DMSQSs have been studied. 

Firstly, we looked into the structural properties of DMSQSs. The EoS of DMSQSs softens with both massive and higher population of fermionic DM and as a result, we obtain massive DMSQS configurations with larger radius for lower values of $m_{\chi}$ and ${k_F}_{\chi}$. This is because the presence of DM makes the star more compact and this compactness is further individually enhanced by heavier and dense population of fermionic DM. Overall, very compact and massive DMSQS configurations, that not only satisfy the constraints from PSR J0740+6620 on the mass-radius relation of compact stars but also the mass of the secondary component of GW190814, are achieved. Many of the DMSQS configurations also satisfy the various recent observational and astrophysical constraints from GW170817, the NICER experiment for PSR J0030+0451, and HESS J1731-347.
\par 
The prominent non-radial $f$-mode frequencies of the DMSQSs are calculated by considering the general relativistic Cowling approximation. By obtaining the $f$-modes as a function of stellar mass, compactness and tidal deformability, we have studied in detail how the presence of DM and its interaction with SQM affect the $f$-mode spectra. We have found that the $f$-mode frequencies of DMSQSs, which are largely populated with massive DM fermions, are higher compared to that of the no-DM scenario (SQSs). We also obtained that the fundamental $f$-mode oscillations are more sensitive to the variations in Fermi momentum of DM ${k_F}_{\chi}$, than to the changes in its mass $m_\chi$. Thus the presence and interaction of DM with SQM has profound impact on the structural and oscillation properties of the DMSQSs. We conclude that the presence of massive DM of higher density resulted in comparatively less massive, but very compact DMSQSs with larger $f$-mode oscillation frequencies. 
An empirical relation between the $f$-modes and the average density $\bar{\rho}=\sqrt{\bar{M}/\bar{R}^3}$ was also determined and we found that the linear fit for DMSQSs differ significantly from the  earlier works. We also estimated relations between the mass-scaled angular frequency $\omega M$ with the compactness $C$ and tidal deformability $\Lambda$. It was observed that $\omega M$ varies universally with $C$ for DMSQSs. 
We also found that for SQS and DMNS, the Cowling approximation has a deviation of about $15-35\%$ when compared with fully relativistic formalism. Further, our investigation reveals that the inclusion of DM reduces the deviation for DMNS, since it makes the star more compact.
Similarly, it is expected that DMSQSs may not introduce any further deviation from the results obtained with fully GR conditions.
In future, we would like to do a general relativistic treatment of the $f$-mode oscillations in strange quark stars with the presence of DM to understand this point better.


\section*{Acknowledgements}
We would like to thank the unknown referee, whose suggestions have improved the quality of the manuscript. We would like to thank S. Shirke for providing the DM EoS data used in Fig. 11 for comparison. L.J.N. acknowledges the Department of Science and Technology, Government of India for the INSPIRE Fellowship. Work of DS is supported by the NRF research Grants (No. 2018R1A5A1025563). Work of AG is supported by the National Research Foundation of Korea, grant funded by Korea Government (MSIT) (RS-2024-00356960). 

\bibliography{QMDMosc}

\end{document}